\documentclass[fleqn,usenatbib]{mnras}
\usepackage{newtxtext,newtxmath,times,verbatim}
\usepackage[T1]{fontenc}
\DeclareRobustCommand{\VAN}[3]{#2}
\let\VANthebibliography\thebibliography
\def\thebibliography{\DeclareRobustCommand{\VAN}[3]{##3}\VANthebibliography}
\usepackage{graphicx}
\usepackage{amsmath}
\usepackage{ulem}
\usepackage{comment}

\title[Nearby analogues of high-redshift dwarfs]{Nearby dwarf galaxies with extreme star formation rates: a window into dwarf-galaxy evolution in the early Universe}

\author[S. Kaviraj et al.]{S. Kaviraj,$^{1}$\thanks{E-mail: s.kaviraj@herts.ac.uk} B. Bichang'a,$^{1}$ I. Lazar,$^{1}$ A. E. Watkins,$^{1}$ G. Martin,$^{2}$ and R. A. Jackson$^{3}$\\
$^{1}$Centre for Astrophysics Research, University of Hertfordshire, Hatfield, AL10 9AB, UK\\
$^{2}$School of Physics and Astronomy, University of Nottingham, University Park, Nottingham NG7 2RD, UK\\
$^{3}$Department of Physics and Astronomy, University of Victoria, Victoria, BC, Canada V8P 5C2}

\begin{document}
\label{firstpage}
\pagerange{\pageref{firstpage}--\pageref{lastpage}}
\maketitle

\begin{abstract}
We study a sample of nearby ($z\sim0.2$) low-luminosity dwarf (10$^{7}$ M$_{\odot}$ < $M_{\rm{\star}}$ < 10$^8$ M$_{\odot}$) galaxies which have extreme {\color{black}(0.1 -- 3 M$_{\odot}$ yr$^{-1}$) star formation rates (SFRs) for this mass regime}, making them plausible analogues of dwarfs at $z\sim5.5$. We compare the properties of these analogues to control samples of `normal' dwarfs, {\color{black}which reside on the star formation main sequence (SFMS) at $z\sim0.2$} and are matched {\color{black}in their stellar mass and redshift distributions to the analogue population}. The analogue and normal populations do not show differences, either in their half-light radii or the projected distances to nodes, filaments and massive galaxies. This suggests that the comparatively extreme SFRs in the analogues are not driven by them being anomalously compact or because they reside in specific environments which might provide a larger gas supply. However, the {\color{black}fractions} of interacting galaxies and those that have early-type morphology are significantly elevated {\color{black}(by factors of $\sim$5.6 and $\sim$9 respectively)} in the analogues compared to the normal population. Extrapolation of the redshift evolution of the star formation main sequence into our mass range of interest appears to underestimate the SFRs of observed dwarfs at $z\sim5.5$. Since current SFMS measurements remain dominated by low and intermediate redshift data (especially at low stellar masses), our study suggests that this underestimation may be driven by interactions (which are more frequent at earlier epochs) boosting the SFRs in the high-redshift dwarf population. Our results are consistent with a picture where higher gas availability, augmented by interactions, drives much of the stellar mass assembly of dwarf galaxies in the early Universe.  
\end{abstract}




\section{Introduction}
\label{sec:intro}

Dwarf galaxies are fundamental to our understanding of galaxy evolution. {\color{black}Not only do dwarfs dominate the galaxy number density at all epochs \citep[e.g.][]{Kaviraj2017,Wright2017,Weaver2023}, their shallow potential wells make them sensitive laboratories for studying many of the processes that shape the evolution of galaxies over cosmic time, such as baryonic feedback \citep[e.g.][]{Martin2019,Manzano-King2019,Koudmani2022,Romano2023,Romano2024}, stellar mass build-up \citep[e.g.][]{Flores-Freitas2024}, mergers and interactions with neighbours \citep[e.g.][]{Martin2021,Jackson2021a,Jackson2021b,Watkins2023} and the formation of small-scale structure such as planes of satellites around massive galaxies \citep[e.g.][]{Revaz2018,Uzeirbegovic2024,Madhani2025}.} Notwithstanding their significance, much of what is known about the dwarf regime is underpinned by studies in our local neighbourhood, e.g. in the Local Group \citep[e.g.][]{Tolstoy2009} or from surveys of dwarf satellites around nearby massive galaxies \citep[e.g.][]{Hunter2012,Madden2013,Duc2015,Geha2017,Venhola2018,Trujillo2021,Poulain2021,Mao2021}. This is driven by the relative faintness of the dwarf population, which makes it difficult to assemble complete unbiased statistical samples outside the nearby Universe using past surveys like the SDSS \citep[e.g.][]{Alam2015} which are wide but relatively shallow \citep{Kaviraj2025}. 

However, new surveys are poised to revolutionise our understanding of dwarf-galaxy evolution, both in the nearby Universe and at high redshift. Surveys such as the Dark Energy Survey \citep[DES;][]{Dey2019} and the Hyper Suprime-Cam Subaru Strategic Program \citep[HSC-SSP;][]{Aihara2018a} have recently enabled the detection and detailed study of dwarfs in low-density environments outside the local neighbourhood \citep[e.g.][]{Tanoglidis2021,Thuruthipilly2023,Lazar2024a,Lazar2024b,Kaviraj2025}. The forthcoming Legacy Survey of Space and Time \citep[LSST;][]{Ivezic2019} will drive another leap in the scope of such work, by increasing the footprints where deep optical imaging is available to roughly half the sky. In a similar vein, the advent of the James Webb Space Telescope \citep[JWST;][]{Gardner2023} is offering novel insights into galaxy evolution in the early Universe, with recent work successfully probing star formation activity and metallicity in high-redshift dwarf galaxies \citep[e.g.][]{Nakajima2023,Boyett2024,Clarke2024,Curti2024,Harshan2024,Duan2024}, down to stellar masses of $M_{\rm{\star}}$ $\sim$ 10$^7$ M$_{\odot}$ at $z\sim5-6$. Nevertheless, studying large, statistical samples of dwarfs down to these stellar masses at high redshift remains challenging, even with the exquisite abilities of the JWST. 

Several studies in the past literature have explored the evolution of high-redshift dwarf galaxies by studying nearby `analogues' i.e. galaxies at low redshift which have similar properties (stellar masses, SFRs etc.) to their counterparts in the early Universe \citep[e.g.][]{Cardamone2009,Yang2017,Lofthouse2017,Bian2018,Izotov2021,Nakajima2022,Shivaei2022}. While low redshift galaxies cannot fully mimic the conditions in which their high-redshift counterparts reside, studying nearby analogues offers an opportunity to study aspects of these galaxies in greater detail than might be possible at high redshift. 

For example, given that they reside in the nearby Universe, higher signal-to-noise observations are possible of analogues, which may enable more accurate estimations of physical parameters like stellar mass, SFR and local environment. Since they reside in the nearby Universe, analogues have larger angular sizes, which aids in the study of structure and morphology. This is particularly true of low-surface-brightness features like tidal tails that signpost recent interactions \citep[e.g.][]{Kaviraj2014b} but which, given the prohibitively large cosmological dimming\footnote{Cosmological dimming in terms of magnitudes scales with redshift as $10 \times \log (1+z)$. The cosmological dimming out to e.g. $z\sim6$ is therefore around 8.5 mag arcsec$^{-2}$.}, are challenging to study outside the nearby Universe \citep[e.g.][]{Martin2022}. Finally, the low redshift of analogues also makes it feasible to study these systems in the rest-frame optical using existing datasets.  

Past dwarf analogue studies have offered interesting insights into the processes that may be important in driving stellar assembly in dwarfs at high redshift. For example, \citet{Cardamone2009} have studied a population of `green pea' galaxies, which are compact, low-metallicity systems with extreme SFRs, making them plausible analogues of relatively luminous dwarfs (10$^{8.5}$ M$_{\odot}$ < $M_{\rm{\star}}$ < 10$^{10}$ M$_{\odot}$) at high redshift. A conclusion of this work, using images from the Hubble Space Telescope (HST), is that the green peas show evidence for star forming clumps and low-surface-brightness features that suggest the presence of recent mergers and interactions. Other studies, which have probed green peas using integral field spectroscopy \citep[e.g.][]{Lofthouse2017} and HI imaging \citep[e.g.][]{Purkayastha2022} have shown that at least some of these systems indeed show strong evidence of interactions \citep[see also][]{Lereste2024}. The importance of interactions in driving star formation in such systems appears to extend to dwarf analogues which have lower stellar masses \citep{Dutta2024}, often called `blueberry' galaxies \citep{Yang2017}. Taken together, these studies hint at galaxy interactions playing a key role in the stellar mass build up of dwarfs in the early Universe. 

The purpose of this work is to perform a {\color{black}homogeneous} statistical study of nearby ($z\sim0.2$) dwarfs in the stellar mass range 10$^{7}$ M$_{\odot}$ < \textit{M}$_{\rm{\star}}$ < 10$^8$ M$_{\odot}$ which have extreme SFRs {\color{black}(0.1 -- 3 M$_{\odot}$ yr$^{-1}$) for this mass regime}, making them plausible analogues of dwarfs at high redshift ($z\sim5.5$). {\color{black}Our choice of $z\sim5.5$ is driven by the fact that our analogue selection is based on extrapolations of the local SFMS out to high redshift, complemented by JWST data of real dwarfs in the early Universe. Currently available SFMS extrapolations extend out to $z\sim5.5$ at which JWST data is also available, making this epoch a natural one at which to base our study.} We combine (1) size measurements, (2) estimates of the projected distances to nodes, filaments and massive galaxies and (3) morphological information derived from visual inspection of deep {\color{black}HSC and HST} images to explore the processes that drive the extreme SFRs in the analogue population. 

The plan for this paper is as follows. In Section \ref{sec:data}, we describe the COSMOS2020 catalogue \citep{Weaver2022} which underpins our study, the selection of our analogue population and a sample of `normal' dwarfs that have typical SFRs at $z\sim0.2$. In the analysis that follows, we compare these two populations to gain insights into the drivers of the star formation activity in the analogues. In Section \ref{sec:properties} we compare the environments (Section \ref{sec:disperse}), sizes (Section \ref{sec:sizes}) and morphologies (Section \ref{sec:morphologies}) of the two populations and speculate on the factors that drive the extreme SFRs in our analogues. We summarise our findings in Section \ref{sec:summary}. 
    

\section{Data}
\label{sec:data}

\subsection{Physical parameters}

Our study is based on physical parameters (photometric redshifts, stellar masses and SFRs) available from the Classic version of the publicly available COSMOS2020 catalogue. COSMOS2020 offers a value-added catalogue of sources in the 2 deg$^2$ COSMOS field, calculated using deep, multi-wavelength UV to mid-infrared photometry, in 40 broad and medium band filters from the following instruments: GALEX \citep{Zamojski2007}, MegaCam/CFHT \citep{Sawicki2019}, ACS/HST \citep{Leauthaud2007}, Subaru/Hyper Suprime-Cam \citep{Aihara2019}, Subaru/Suprime-Cam \citep{Taniguchi2007,Taniguchi2015}, VIRCAM/VISTA \citep{McCracken2012} and IRAC/Spitzer \citep{Ashby2013,Steinhardt2014,Ashby2015,Ashby2018}. A particular strength of this dataset is the use of optical ($grizy$) data for object detection, from the `Ultra-deep' layer of the HSC-SSP \citep{Aihara2019}, which has a point source depth of $r\sim28$ mag. As a comparison, the detection limit of HSC-SSP Ultra-deep is $\sim$5 mag fainter than standard-depth imaging from the SDSS and $\sim$10 mag fainter than the magnitude limit ($r\sim18$ mag) of the {\color{black}SDSS spectroscopic main galaxy sample}.  

Fluxes are extracted within circular apertures after the survey images are homogenized to a common point-spread function. The \textsc{SExtractor} and \textsc{IRACLEAN} codes are used to process the UV/optical and infrared photometry respectively. {\color{black}Physical parameters are then calculated by applying the \textsc{LePhare} SED-fitting algorithm \citep{Arnouts2002,Ilbert2006} to this UV -- mid infrared photometry. The template library, described in \citet{Ilbert2009}, spans templates from all galaxy morphological types \citep{Polletta2007}, blue star-forming galaxy models from \citet{Bruzual2003}, exponentially declining star formation histories to improve the photometric redshifts of quiescent galaxies \citep[][]{Onodera2012} and templates that account for AGN \citep[e.g.][]{Salvato2009,Salvato2011}. Extinction is included as a free parameter, with a flat prior in the reddening ($E_{\rm B-V}$ < 0.5). The attenuation curves considered in the SED fitting are taken from \citet{Calzetti2000}, \citet{Prevot1984} and two variants of the Calzetti law which include the bump at
2175 \AA \: \citep{Fitzpatrick1986} with two different
amplitudes. Emission lines are added using the relation between the UV luminosity and [O II] emission-line flux, following \citet{Ilbert2009}.} The accuracies of the resultant photometric redshifts are better than 1 and 4 per cent for bright ($i<22.5$ mag) and faint ($25<i<27$ mag) galaxies respectively. We direct readers to \citet{Weaver2022} for more details of the construction of the catalogue. 

\begin{figure}
\center
\includegraphics[width=\columnwidth]{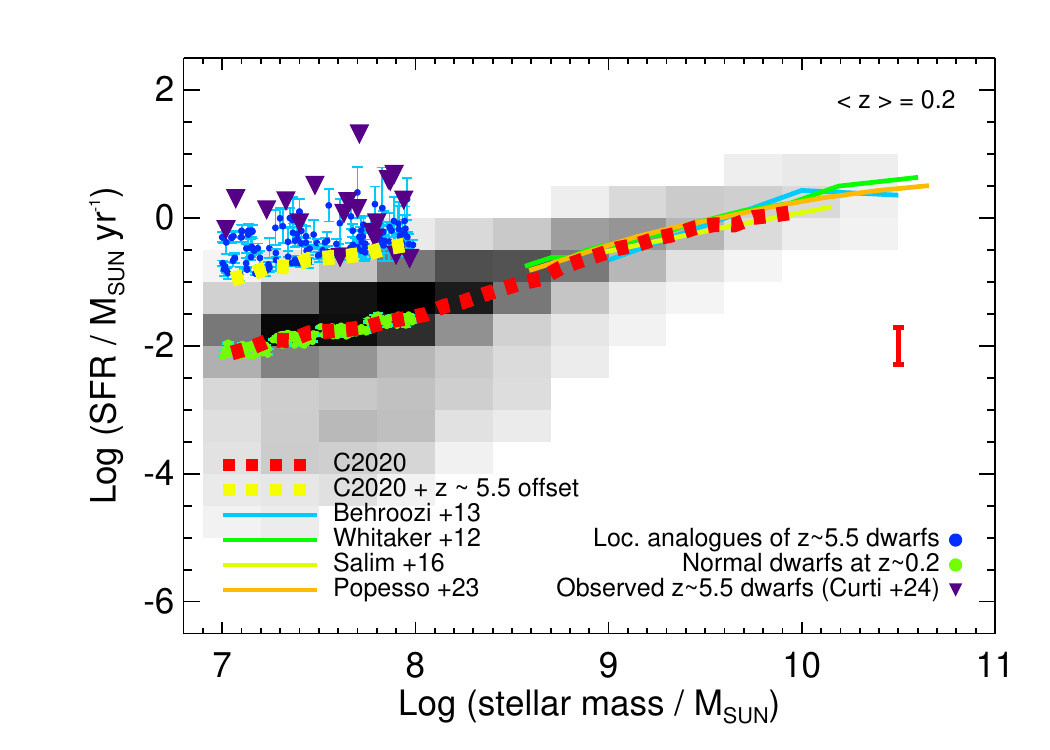}
\caption{The SFMS for galaxies at $z<0.3$ in the COSMOS2020 catalogue. The median redshift of this population is $z\sim0.2$. {\color{black}The heatmap shows all COSMOS2020 galaxies which reside at $z<0.3$}, the thick dotted red line shows the ridgeline of the SFMS at $z\sim0.2$ and the thick dotted yellow line shows the ridgeline of the SFMS at $z\sim0.2$ plus the $z\sim5.5$ offset calculated using P23 (see text in Section \ref{sec:selection} for details). {\color{black}Other SFMS measurements at $z\sim0.2$ from the literature \citep{Whitaker2012,Behroozi2013,Salim2016}, which agree well with the COSMOS2020 $z\sim0.2$ ridgeline, are shown using the thin solid lines. The average scatter in these SFMS measurements is shown using the red error bar.} The blue and green circles show $z\sim5.5$ analogues and the population of normal dwarfs respectively (see Section \ref{sec:selection} for how these populations are defined). SFR uncertainties are shown for individual galaxies. The purple triangles indicate SFR measurements of galaxies which have stellar masses in the range 10$^{7}$ M$_{\odot}$ < $M_{\rm{\star}}$ < 10$^{8}$ M$_{\odot}$ and redshifts in the range $4.5<z<6$ from \citet{Curti2024}.}
\label{fig:delta}
\end{figure}

To construct an initial sample of dwarfs, we select objects that are (1) classified as galaxies by \textsc{LePHARE} (\texttt{type} = 0 in the COSMOS2020 catalogue), (2) classified as `extended' (i.e. galaxies) by the HSC pipeline in the HSC $g$, $r$, $i$ and $z$ filters, (3) have stellar masses in the range 10$^{7}$ M$_{\odot}$ < $M_{\rm{\star}}$ < 10$^{8}$ M$_{\odot}$ and redshifts in the range $z<0.3$, (4)
have $u$-band as well as near and mid-infrared photometry (since a wide wavelength baseline aids the accuracy of the parameter estimation, see e.g. \citet{Ilbert2006}) and (5) lie outside masked regions such as bright star masks and image edges (\texttt{flag\_combined} = 0 in the COSMOS2020 catalogue). This produces a sample of $\sim$6400 dwarf galaxies with a median redshift of $z\sim0.2$. As described below, we use this sample to select both a population of high-redshift dwarf analogues and a sample of `normal' dwarfs in the nearby Universe to which the analogues can be compared, in order to investigate the processes that drive their extreme SFRs.

  
\subsection{Selection of candidate samples of high-redshift analogues and normal dwarfs}
\label{sec:selection}

We use the dwarf galaxy sample described above to select two populations. The first is a sample of dwarfs which have, compared to dwarf galaxies at low redshift, extreme SFRs that make them plausible analogues of the dwarf population at high redshift ($z\sim5.5$). The second is a sample of `normal' dwarfs which have typical SFR properties in the nearby Universe. {\color{black}These are defined as systems which lie between the 45\textsuperscript{th} and 55\textsuperscript{th} percentile values of the SFR distribution at a given stellar mass i.e. they straddle the evolving median value of the SFR distribution as a function of stellar mass in the nearby Universe (e.g. at $M_{\rm \star}$ $\sim$ 10$^{7}$ M$_{\odot}$ the log of the median SFR is around -2).} In Section \ref{sec:properties}, we compare the analogues to control samples, constructed from the normal population, which are matched to the analogues in stellar mass and redshift, in order to explore the source of the extreme SFRs in our analogues. 

The identification of analogues ideally requires statistical samples of SFR measurements of dwarf galaxies at high redshift in our mass range of interest (10$^{7}$ M$_{\odot}$ < $M_{\rm{\star}}$ < 10$^{8}$ M$_{\odot}$). {\color{black}With the advent of JWST, small samples of galaxies in this mass range have become available at high redshift \citep[e.g.][]{Curti2024}}. To select our analogues, we use both measurements of the SFMS at $z\sim5.5$ and measured dwarf SFRs at high redshift. {\color{black}As noted before, our choice of using $z\sim5.5$ as our definition of the high redshift Universe results from the fact that currently available measurements of the SFMS extend out to this epoch and JWST data is also available at these redshifts.}  

Figure \ref{fig:delta} shows SFR vs stellar mass for galaxies at $z<0.3$ in the COSMOS2020 catalogue. Recall that the median redshift of this population is $z\sim0.2$. {\color{black}The heatmap shows all COSMOS2020 galaxies which reside at $z<0.3$}, while the thick dotted red line shows a running median, which we use to define the ridgeline of the SFMS at $z\sim0.2$. Other SFMS measurements at $z\sim0.2$ from the literature (see legend), which agree well with the COSMOS2020 ridgeline, are shown using the solid lines. To select our analogue population we proceed as follows. \citet[][P23 hereafter]{Popesso2023} have measured the evolution of the SFMS out to high redshift and demonstrate that the offset between the local SFMS and those at progressively higher redshifts does not depend strongly on stellar mass at $M_{\rm \star}$ < 10$^{9.5}$ M$_{\odot}$. Therefore, we first calculate the P23 SFMS offset, measured at 10$^{8.8}$ M$_{\odot}$, between the local SFMS and that at $z\sim5.5$ (the highest redshift out to which P23 offsets are available). We then add this offset {\color{black}(which is around 1 dex)} to the COSMOS2020 ridgeline at $z\sim0.2$ to estimate the extrapolated position of the SFMS at $z\sim5.5$ in our mass range of interest. This is shown using the thick dotted yellow line in Figure \ref{fig:delta}. The purple triangles indicate SFR measurements of observed galaxies which have stellar masses in the range 10$^{7}$ M$_{\odot}$ < $M_{\rm{\star}}$ < 10$^{8}$ M$_{\odot}$ and redshifts in the range $4.5<z<6$ from \citet{Curti2024}. 

\begin{figure*}
\center
\includegraphics[width=2\columnwidth]{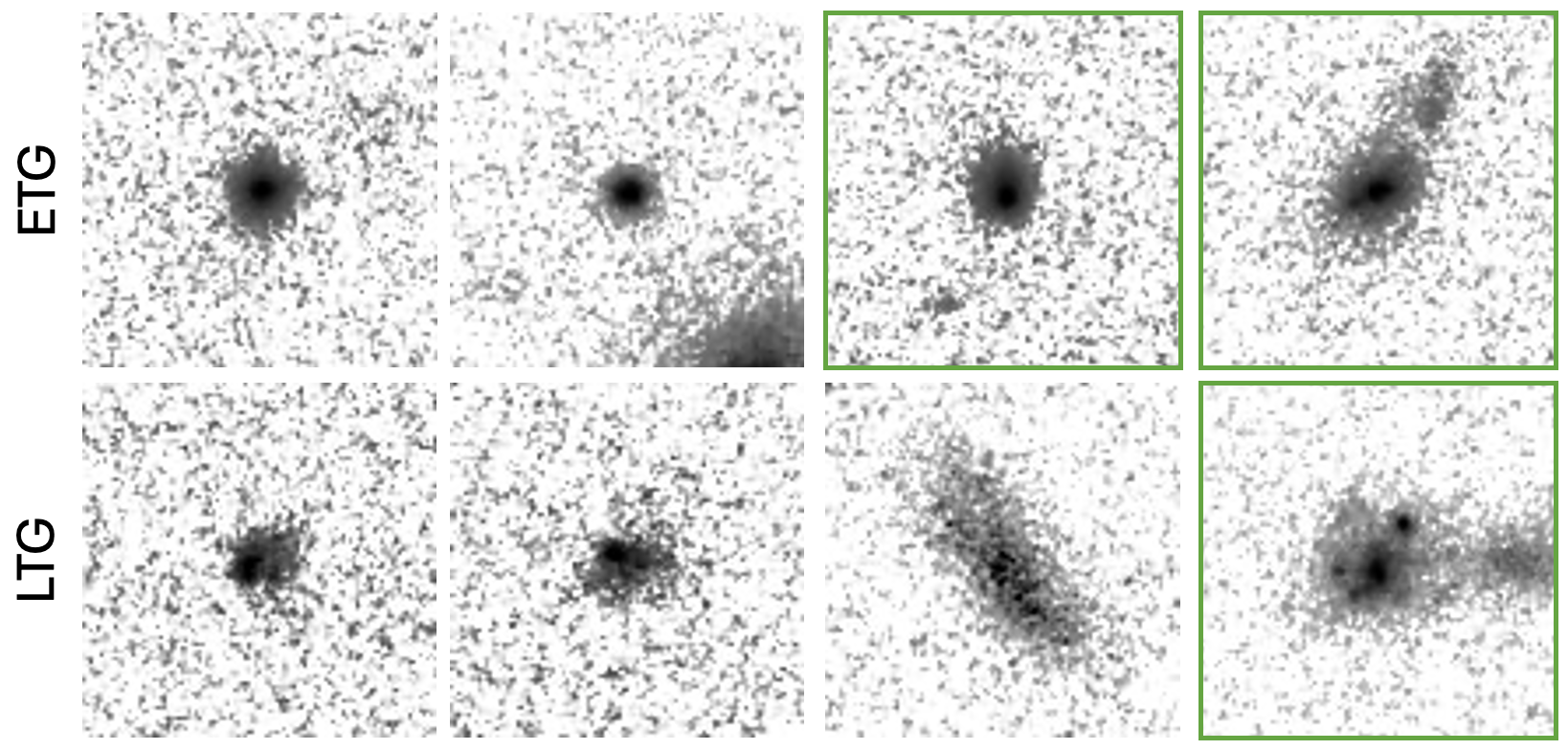}
\caption{HST F814W images of a random sample of our dwarf galaxies. The first row shows examples of galaxies classified as ETGs, while the bottom row shows LTGs. {\color{black}The panels with a green border show examples of systems that are flagged as interacting.} The ETG in column 3 shows an internal asymmetry, while the ETG in column 4 shows a tidal feature to the north east of the galaxy. The LTG in column 4 similarly shows a tidal feature to the east of the galaxy. Each image has a size of 3 arcsec on a side. At the median redshift of our dwarf sample ($z\sim0.2$), this corresponds to a physical size of around 10 kpc.}
\label{fig:example_images}
\end{figure*}

The measured SFRs of high-redshift dwarfs lie almost exclusively \textit{above} the thick dotted yellow line, as opposed to straddling its position in the SFR vs stellar mass plane. This suggests that the extrapolation of the P23 SFMS -- which is underpinned by relatively massive galaxies at low and intermediate redshift -- into our mass range of interest, may underestimate the SFRs of dwarfs at high redshift \citep[as has been noted already in previous work, e.g.][]{Curti2024}\footnote{{\color{black}We note that a potential reason for this discrepancy could be that the JADES survey \citep{Eisenstein2023}, which underpins the \citet{Curti2024}, may preferentially detect dwarfs with very high SFRs. This is not possible to confirm without a deeper survey that is capable of observing more quiescent galaxies. However, it is worth noting that the spread ($\sim$1 dex) in the observed SFRs of high-redshift dwarfs does not decrease in our mass range of interest compared to slightly more massive galaxies. Furthermore, the fraction of galaxies for which spectroscopic redshifts can be  successfully measured remains stable out to $z\sim6$ \citep[e.g.][]{Bunker2024}. Together this suggests that JADES is unlikely to be significantly skewed towards the objects with highest SFRs at $z\sim6$}}. We return to this point later during our study of the factors that drive the extreme SFRs in the analogue population. Given this comparison, we select an initial sample of analogue candidates as galaxies which, at a given stellar mass, have SFRs equal to or greater than the locus traced by the thick dotted yellow line. The final sample of analogues is selected after visually inspecting the HSC and HST images of each galaxy to remove cases where more than one source may be contributing to the HSC photometry, as described in the next section.   

Finally, our population of normal dwarfs, which is selected to lie around the ridgeline of the $z\sim0.2$ SFMS, as described above, is shown using the green circles along with their error bars. As with the analogues, we select an initial sample of normal candidates, from which we then select a final sample after the visual inspection described in the next section. 


\subsection{Visual inspection of HSC and HST images - removal of blended sources and morphological classification}
\label{sec:visualinspection}

We visually inspect both the HSC $gri$ colour composite and HST F814W ($I$-band) images \citep{Koekemoer2007,Massey2010} of each analogue and normal galaxy. {\color{black}The HSC and HST images have 5$\sigma$ point-source depths of 27.8 and 28 mag and angular resolutions of $\sim$0.6 \citep{Aihara2018b} and $\sim$0.05 arcsec \citep{Koekemoer2007} respectively.} The visual inspection, performed by one expert classifier (SK), has two purposes. {\color{black}First, as noted above, we use the higher resolution HST images to identify cases where the HSC photometry may have contributions from two separate sources - we find 8 objects matching this description\footnote{{\color{black}The fraction of such blended sources is similar ($\sim$6 per cent) in the analogue and normal populations.}}.} In such cases, the HSC centroid lies between the two HST objects that have been blended in the HSC photometry. Appendix \ref{sec:blended_sources} shows an example of such systems, which we remove from our analysis. Our final samples of analogue and normal dwarfs contain 116 and 590 galaxies respectively. The blue and green circles in Figure \ref{fig:delta} show the final analogue and normal populations that underpin our study, with individual SFR error bars also shown for each galaxy.  

The second purpose of the visual inspection is to perform morphological classification of our dwarf galaxies using their HST images. {\color{black}Visual inspection offers the most accurate route to the classification of galaxy morphologies and the identification of interacting systems \citep[e.g.][]{Kaviraj2014b}, against which automated methods which use quantitative parameters are often calibrated \citep[e.g.][]{Lazar2024a}}. We classify our galaxies into two broad morphological classes: early-type galaxies (ETGs) and late-type galaxies (LTGs), following the traditional definitions of these morphological types. {\color{black}ETGs exhibit central light concentrations which are surrounded by smooth light distributions, while galaxies that do not exhibit these characteristics are classified as LTGs.} We also flag interacting systems that show evidence of an ongoing or recent interaction e.g. tidal features, internal asymmetries or tidal bridges due to an ongoing merger with another galaxy. 

Note that none of the 8 objects that were removed from our analysis because the HSC photometry may have contributions from two separate sources show any evidence of tidal features or tidal bridges between the two objects suggesting that they are interacting. The removal of these objects therefore has no impact on the fraction of interacting galaxies in our sample. Finally, we note that our interacting galaxies are, by construction, those that are already in the process of interacting i.e. they are not `close pairs', which are not possible to identify without spectroscopic redshifts. 

Figure \ref{fig:example_images} shows HST images of a random sample of our dwarf galaxies. The first row shows examples of galaxies classified as ETGs, while the bottom row shows LTGs. {\color{black}The panels with a green border show examples of systems that are flagged as interacting.} The ETG in column 3 shows an internal asymmetry, while the ETG in column 4 shows a tidal feature to the north east of the galaxy. The LTG in column 4 similarly shows a tidal feature to the east of the galaxy. Each image has a size of 3 arcsec on a side. At the median redshift of our dwarf sample ($z\sim0.2$), this corresponds to a physical size of around 10 kpc.


\subsection{Estimation of environmental parameters}

We measure the distances of our dwarfs from nodes, filaments and massive ($M_{\star}$ > 10$^{10}$ M$_{\odot}$) galaxies using the DisPerSE \citep{Sousbie2011} structure-finding algorithm. Our methodology follows \citet{Lazar2023} and \citet{Bichanga2024}, who have performed a similar density analysis using the COSMOS2020 catalogue. DisPerSE creates a density map using Delaunay tessellations, calculated using the positions of galaxies \citep{Schaap2000}. Stationary i.e. critical points in the density map (minima, maxima and saddles) correspond to voids, nodes and the centres of filaments in the cosmic web respectively. Segments are then used to connect the saddle points with the local maxima, forming a set of ridges that describes the network of filaments that define the cosmic web. The properties of the filament network thus created are determined by a `persistence' parameter, which sets a threshold value for defining pairs of critical points within the density map. {\color{black}Only critical pairs with Poisson probabilities above \textit{N}$\sigma$ from the mean are retained.} We use a persistence value of 2 in our study, which removes ridges close to the noise level, where structures could be spurious.

We use massive ($M_{\star}$ > 10$^{10}$ M$_{\odot}$) galaxies to construct our density maps, as they have the smallest redshift errors {\color{black}($\delta z \sim0.008$ at $z\sim0.2$)} and dominate the local gravitational potential wells. 
The accuracy of the COSMOS2020 redshifts enables us to employ well-defined and relatively narrow ($\delta z\sim0.02$) redshift slices to build our density maps. When constructing each density map, individual galaxies are weighted by the area under their redshift probability density function that is contained within the slice in question. This ensures that the photometric redshifts of massive galaxies, albeit very accurate in COSMOS2020, are propagated through the construction of the density maps. 


\section{Comparison of the properties of analogue and normal dwarfs}
\label{sec:properties}

We proceed by comparing the properties of the analogue and normal populations - local environment, galaxy sizes, the morphological mix and the presence of interactions. In all cases, we compare the analogue population to a control sample drawn from the normal dwarfs, which has an identical number of galaxies and is matched to the analogues in its distributions of stellar mass and redshift.  


\subsection{Distances from nodes, filaments and massive galaxies}
\label{sec:disperse}

\begin{figure}
\center
\includegraphics[width=\columnwidth]{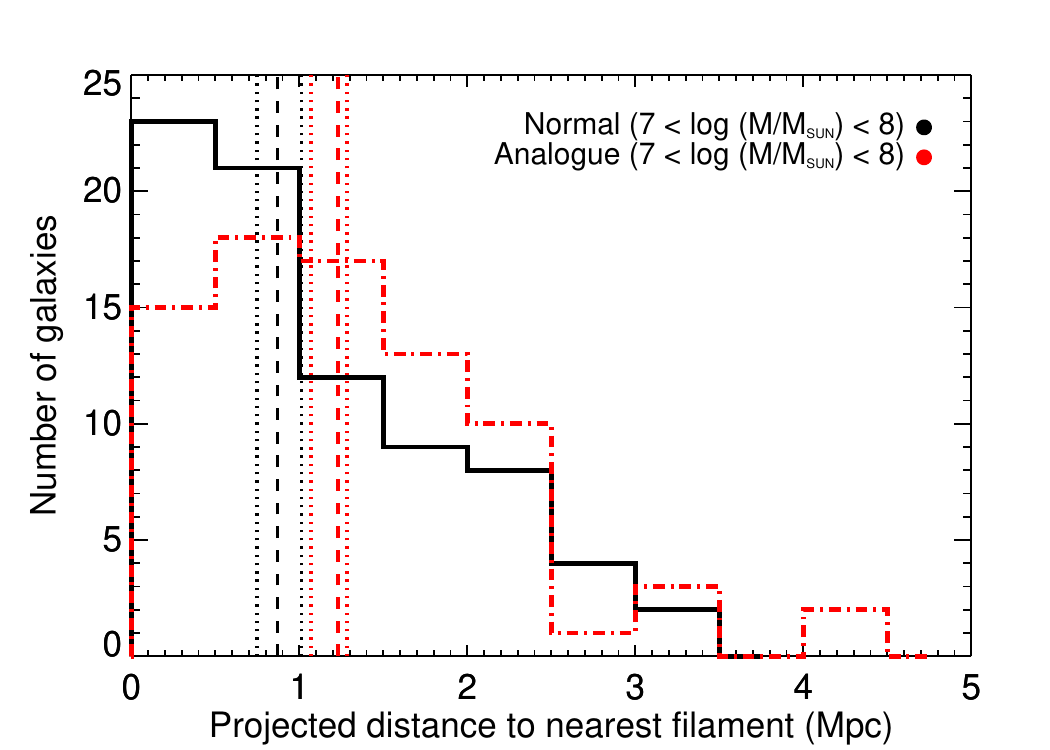}\\
\includegraphics[width=\columnwidth]{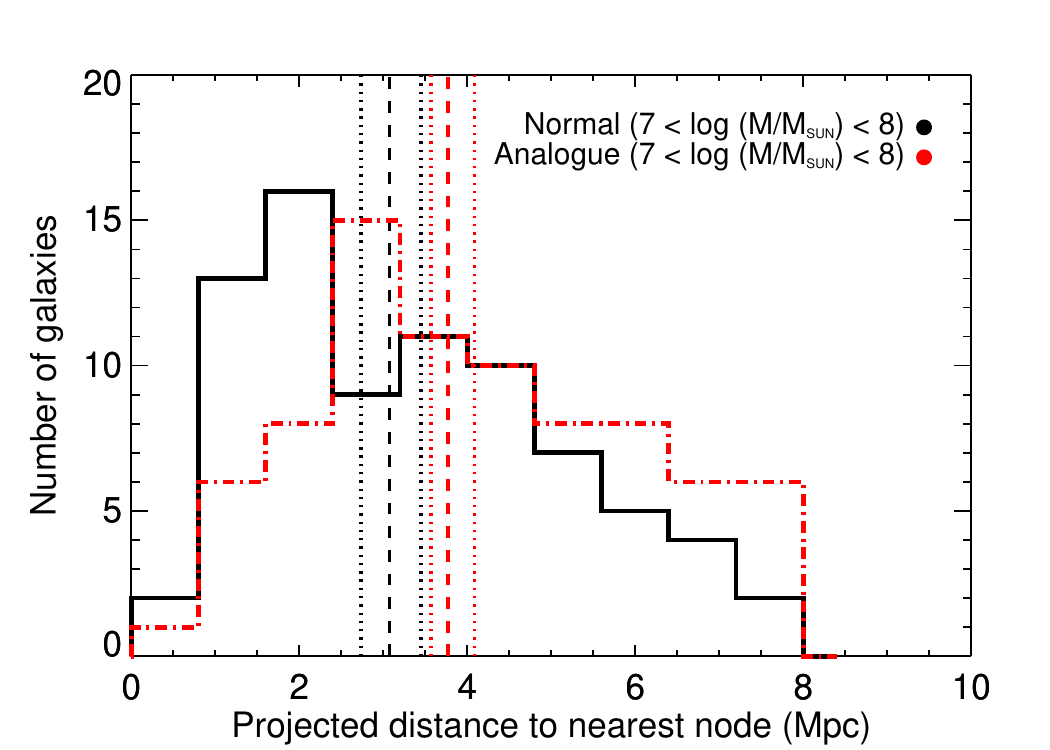}\\
\includegraphics[width=\columnwidth]{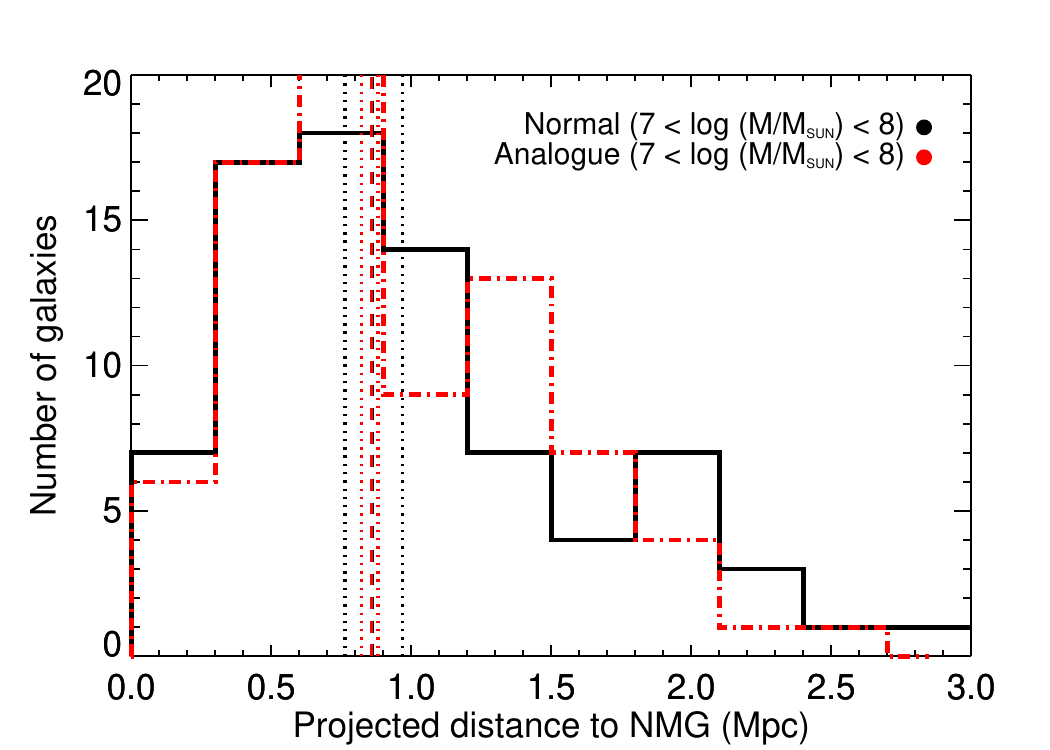}
\caption{Distributions of projected distances from the nearest filaments (top), nodes (middle) and massive ($M_{\star}$ > 10$^{10}$ M$_{\odot}$) galaxies (bottom) for the analogue (red) and normal (black) populations. {\color{black}NMG = nearest massive galaxy.} Median values and their uncertainties are shown using the dashed and dotted lines respectively. {\color{black}These are calculated using bootstrapping, where the distribution is resampled with replacement 1000 times. The median for each resampled distribution is calculated and the standard deviation of the resultant distribution of 1000 median values is used as the error on the median of the original distribution.}}
\label{fig:environment}
\end{figure}

We begin, in Figure \ref{fig:environment}, by comparing the distributions of projected distances from the nearest filaments, nodes and massive ($M_{\star}$ > 10$^{10}$ M$_{\odot}$) galaxies for the analogue and normal populations. Median values and their uncertainties are shown using the dashed and dotted lines respectively. {\color{black}Compared to the normal population}, the analogues reside slightly further away from filaments and nodes but, given the errors on the medians, show no difference in the median distance to the nearest massive galaxies. KS tests for the two populations yield {\color{black}p-values} of 0.32, 0.06 and 0.52 for the projected distances to filaments, nodes and massive galaxies respectively. {\color{black}Recall that the null hypothesis in the KS test, i.e. that the two samples are drawn from the same distribution, is typically rejected if the p-value is lower than 0.05.} While the median values indicate that the environments of the analogue and normal dwarfs are slightly different, the KS tests suggest that these differences are relatively small (except possibly for the distances to nodes, where the p-value is close to 5 per cent). These minor differences suggest that the large discrepancy in the SFRs between the analogue and normal populations is not driven by differences in local environment. 

\begin{figure}
\center
\includegraphics[width=\columnwidth]{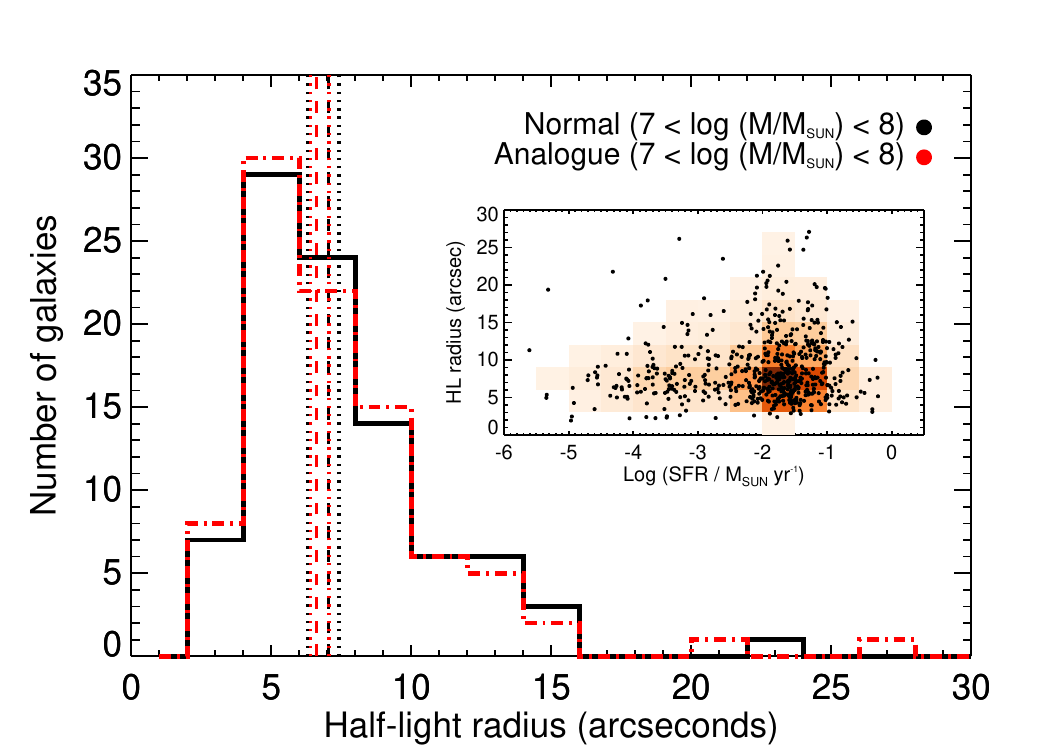}
\caption{Distributions of half-light radii of the analogue (red) and normal (black) populations. Median values and their uncertainties (calculated using bootstrapping) are shown using the dashed and dotted lines respectively. The inset shows the half-light radius plotted against the SFR. The heatmap shows all dwarfs which have stellar masses in the range 10$^{7}$ M$_{\odot}$ < $M_{\rm{\star}}$ < 10$^{8}$ M$_{\odot}$ and redshifts in the range $z<0.3$. A random 10 per cent of galaxies are shown overplotted using the black dots.}
\label{fig:sizes}
\end{figure}


\subsection{Galaxy size} 
\label{sec:sizes}

Next we consider the sizes of the analogue and normal populations. Figure \ref{fig:sizes} compares the distributions of half-light radii, calculated in the $I$-band \citep{Leauthaud2007}, of these two samples. {\color{black}The half-light radius is calculated using the curve of growth of the flux and defined as the radius that contains half of the light from the galaxy.} The median values are indistinguishable within the uncertainties and a KS test yields a p-value of 0.1 suggesting that the distributions are similar. The inset, which shows the half-light radii plotted against the SFRs of all dwarf galaxies in our mass range of interest, indicates that galaxy size generally does not show a correlation with SFR, in agreement with results in the literature at higher stellar masses \citep[e.g.][]{Whitaker2017,Lin2020}. The lack of a correlation between SFR and size across the general dwarf population explains why the sizes of the analogues and normal samples are similar even though they have very different SFRs. In summary, the differences in the SFRs between the analogue and normal populations are unlikely to be driven by the analogues having different sizes (e.g. due to being more compact). 


\subsection{Morphology and the presence of interactions}
\label{sec:morphologies}

Table \ref{tab:morphologies} summarises the fractions of galaxies classified as ETG and interacting in the analogue and normal populations. {\color{black}The uncertainties in the fractions are estimated following \citet{Cameron2011} who calculate accurate Bayesian binomial confidence intervals using the quantiles of the beta distribution. These are more accurate than simpler techniques, like using the normal approximation, which tends to misrepresent the statistical uncertainty under the sampling conditions typically encountered in astronomical surveys.} Both the interacting and ETG fractions are significantly elevated (by several factors) in the analogues, compared to their normal counterparts. 

Given that interactions create and consolidate dispersion-dominated components in galaxies \citep[e.g.][]{Martin2018_sph}, one expects a galaxy population with a higher incidence of interactions to also show an elevated ETG fraction, as appears to be the case here. {\color{black}While the larger interaction fractions in the analogues may be somewhat surprising given the similarities in galaxy environments, it is worth noting that dwarfs have shallow potential wells which make them more susceptible to being morphologically affected by tidal perturbations from nearby objects, so that higher density environments may not be needed for generating the interaction signatures probed here \citep[e.g.][]{Jackson2021b}. It is interesting to note that the median SFR of interacting galaxies is elevated by $\sim$29 per cent compared to those of non-interacting galaxies. In a similar vein the median SFR of galaxies flagged as ETGs is elevated by $\sim$55 per cent compared to those that are LTGs. Taken together, the higher interacting and ETG fractions and the elevations in the SFR in the interacting and ETG populations suggest that interactions play a significant role in driving the extreme SFRs seen in the analogue population.}  

\begin{table}
\begin{center}
\begin{tabular}{ l | c | c | c}
\hline
& Analogue  & Normal & Ratio (Ana./Norm.)\\
\hline
\hline
Interacting fraction & $0.36 \pm 0.05$ & $0.04 \pm 0.02$ & 9\\
ETG fraction         & $0.28 \pm 0.03$ & $0.05 \pm 0.02$ & 5.6\\
\end{tabular}
\caption{The fractions of galaxies that are interacting (second row) and those that have early-type morphology (third row) in the analogue (second column) and normal (third column) populations. {\color{black}Note that any galaxy which does not have early-type morphology is classed as having late-type morphology.} The ratio between the analogue and normal fractions is shown in the last column. The uncertainties are calculated following \citet{Cameron2011}.}
\label{tab:morphologies}
\end{center}
\end{table}

\subsection{What drives the extreme star formation rates in the analogue population?}

The analysis above indicates that the vastly different SFRs of the analogue and normal populations is unlikely to be driven either by a difference in local environment or galaxy size. However, the analogue and normal dwarfs do show significantly different interacting and ETG fractions, suggesting that interactions may play a role in the SFR divergence observed between these two populations. 

\begin{figure}
\center
\includegraphics[width=\columnwidth]{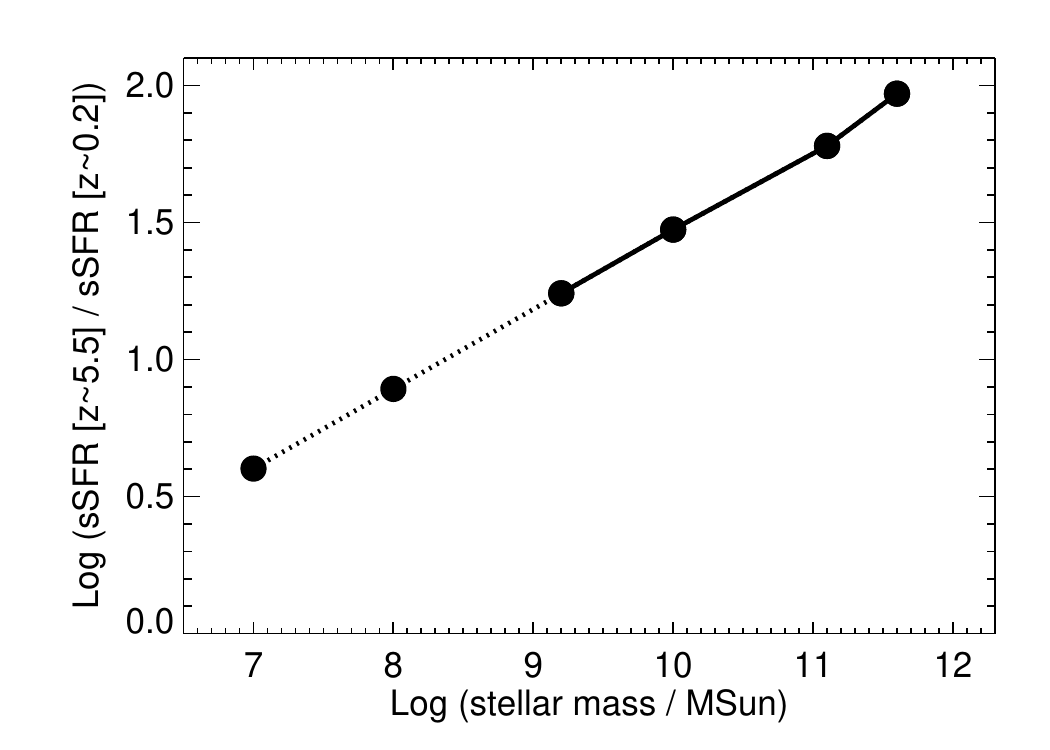}
\caption{The ratio of the sSFR between $z\sim5.5$ and $z\sim0.2$ as a function of stellar mass, from \citet{Liu2019}. While this study does not extend past $M_{\rm \star} \sim 10^{9.2} {\rm \: M_{\odot}}$, the log of the sSFR ratio decreases almost linearly with the log of the stellar mass. Extrapolation of this behaviour into our mass range of interest is shown using the dotted line.}
\label{fig:ssfr_ratio}
\end{figure}

In this context, recall that an extrapolation of the redshift evolution of the P23 SFMS measurements (which are largely based on relatively massive galaxies at low and intermediate redshift) into our mass range of interest appears to underestimate the measured SFRs of dwarfs at $z\sim5.5$. In a similar vein, \citet{Liu2019} have estimated molecular gas fractions ($\mu_{\rm mol gas} \equiv M_{\rm molgas}/M_{\rm \star}$) and gas depletion timescales ($\tau_{\rm molgas} \equiv M_{\rm molgas}/{\rm SFR}$) in relatively massive galaxies ($M_{\rm \star} > 10^{9.2} {\rm \: M_{\odot}}$) out to $z\sim6$. A caveat here is that their fits to the redshift evolution of $\mu_{\rm molgas}$ and $\tau_{\rm molgas}$, particularly at lower stellar masses, relies mainly on data in the nearby Universe. Since $\mu_{\rm mol gas}/\tau_{\rm mol gas}$ yields the specific SFR (sSFR), these results offer a way of gauging the expected evolution of star formation activity due to the changes in the gas conditions between the local Universe and high redshift. 

\citet{Liu2019} show that this redshift evolution in the star formation activity is stellar mass dependent. Figure \ref{fig:ssfr_ratio} shows the ratio of the sSFR (i.e. $\mu_{\rm mol gas}/\tau_{\rm mol gas}$) for galaxies on the SFMS between $z\sim5.5$ and $z\sim0.2$, as a function of stellar mass from \citet{Liu2019}. While this study does not extend past $M_{\rm \star} \sim 10^{9.2} {\rm \: M_{\odot}}$, the log of the sSFR ratio decreases almost linearly with the log of the stellar mass. Extrapolation of this behaviour into our mass range of interest is shown using the dotted line. At $M_{\rm \star} \sim 10^{7} {\rm \: M_{\odot}}$ and $M_{\rm \star} \sim 10^{8} {\rm \: M_{\odot}}$ the sSFR ratios are $\sim$4 and $\sim$8 respectively. Given that the observed SFRs of dwarfs in our mass range of interest are at least a factor of 14 (1.16 dex) higher at $z\sim5.5$ than at $z\sim0.2$, a simple extrapolation of the gas conditions out to high redshift and down to low stellar mass also appears to underpredict the evolution in the star formation activity.  

Given the significantly elevated interacting and ETG fractions found in the analogue population, we speculate that the underestimates described above may be driven, at least partly, by interaction-driven enhancement of star formation in high-redshift dwarf galaxies. The star-formation and gas-mass calibrations in the current literature, particularly at low stellar masses, remain dominated by galaxies at low redshift, where interactions are less frequent \citep[e.g.][]{Conselice2014}. However, it is well established that interactions enhance star formation in dwarf galaxies \citep[e.g.][]{Martin2022,Lazar2024a} in which interactions typically boost star formation across the entire body of the galaxy \citep{Lazar2024b}. For this reason, extrapolating the star formation and gas properties of nearby dwarfs out to high redshift may underestimate the star formation activity in the high-redshift dwarf population. We suggest, therefore, that while gas availability is higher in the early Universe, the stellar assembly in high-redshift dwarfs is catalysed by interactions. {\color{black}Indeed the shallow potential wells of dwarfs may make the impact of interactions greater in these objects than in the massive-galaxy regime, where interactions may not play as significant a role in galaxy assembly at high redshift \citep[e.g.][]{Duncan2019,Romano2021,Dalmasso2024}.

Finally, our conclusions about the role of interactions is driven by the fact that this is where the analogue and normal populations differ the most, while other properties (e.g. galaxy size and environment) seem to be similar. However, a potential caveat is that our study necessarily uses extrapolations of low redshift data which may not accurately represent the conditions in the high redshift Universe, which is worth bearing in mind when considering our conclusions.} 


\section{Summary}
\label{sec:summary}

We have studied a sample of nearby ($z\sim0.2$) low-luminosity dwarf (10$^{7}$ M$_{\odot}$ < \textit{M}$_{\rm{\star}}$ < 10$^8$ M$_{\odot}$) galaxies with extreme SFRs, which make them plausible analogues of dwarfs at $z\sim5.5$. We have first constructed control samples, which are drawn from a population of `normal' dwarfs with typical SFRs at $z\sim0.2$, and are matched in their stellar mass and redshift distributions to the analogue population. We have then compared the properties of the analogues to the normal dwarfs (via the control samples) to explore the processes and factors that are likely to drive the extreme SFRs in the analogues. Our main conclusions are as follows: 

\begin{itemize}

    \item The analogue and normal populations do not show strong differences in their distances to nodes, filaments and massive galaxies, suggesting that the analogue SFRs are not the result of residing in a specific environment which may provide a larger gas supply. 

    \item The analogue and normal populations also do not show differences in their distributions of half-light radii, indicating that the extreme SFRs in the analogues are not driven by a difference in size (e.g. due to the analogues being anomalously compact).

    \item However, the fraction of interacting galaxies and those that are ETGs are significantly elevated (by several factors) in the analogues compared to the normal galaxies, suggesting that interactions play an important role in driving the extreme SFRs.  

    \item Extrapolation of the redshift evolution of the SFMS into our mass range of interest underestimates the SFRs of observed dwarfs at $z\sim5.5$. Since current SFMS measurements remain dominated by data at low and intermediate redshift, we suggest that this underestimation could be driven by interactions, which are more frequent at earlier epochs, boosting the SFRs in the high-redshift dwarf population.

    \item Our results suggest that, while gas availability is higher in the early Universe, the stellar mass assembly of high-redshift dwarfs is catalysed by interactions. 

\end{itemize}


\section*{Acknowledgements}

We thank the referee for many constructive comments that helped us to significantly improve the quality of the original manuscript. SK, IL and AEW acknowledge support from the STFC (grant numbers ST/Y001257/1, ST/S00615X/1 and ST/X001318/1). SK also acknowledges a Senior Research Fellowship from Worcester College Oxford. BB and IL acknowledge PhD studentships funded by the Centre for Astrophysics Research at the University of Hertfordshire. 

The Hyper Suprime-Cam (HSC) collaboration includes the astronomical communities of Japan and Taiwan, and Princeton University. The HSC instrumentation and software were developed by the National Astronomical Observatory of Japan (NAOJ), the Kavli Institute for the Physics and Mathematics of the Universe (Kavli IPMU), the University of Tokyo, the High Energy Accelerator Research Organization (KEK), the Academia Sinica Institute for Astronomy and Astrophysics in Taiwan (ASIAA), and Princeton University. Funding was contributed by the FIRST program from the Japanese Cabinet Office, the Ministry of Education, Culture, Sports, Science and Technology (MEXT), the Japan Society for the Promotion of Science (JSPS), Japan Science and Technology Agency (JST), the Toray Science Foundation, NAOJ, Kavli IPMU, KEK, ASIAA, and Princeton University. This paper makes use of software developed for Vera C. Rubin Observatory. 

This paper is based on data collected at the Subaru Telescope and retrieved from the HSC data archive system, which is operated by the Subaru Telescope and Astronomy Data Center (ADC) at NAOJ. Data analysis was in part carried out with the cooperation of Center for Computational Astrophysics (CfCA), NAOJ. We are honored and grateful for the opportunity of observing the Universe from Maunakea, which has the cultural, historical and natural significance in Hawaii. 

For the purpose of open access, the authors have applied a Creative Commons Attribution (CC BY) licence to any Author Accepted Manuscript version arising from this submission. 


\section*{Data Availability}
The physical parameters used in this study available in  \citet{Weaver2022}. The density parameters were produced using the DisPerSE algorithm \citep{Sousbie2011}. 
 

\appendix

\section{Identification of blended sources in HSC images via visual inspection}
\label{sec:blended_sources}

\begin{figure}
\center
\includegraphics[width=0.8\columnwidth]{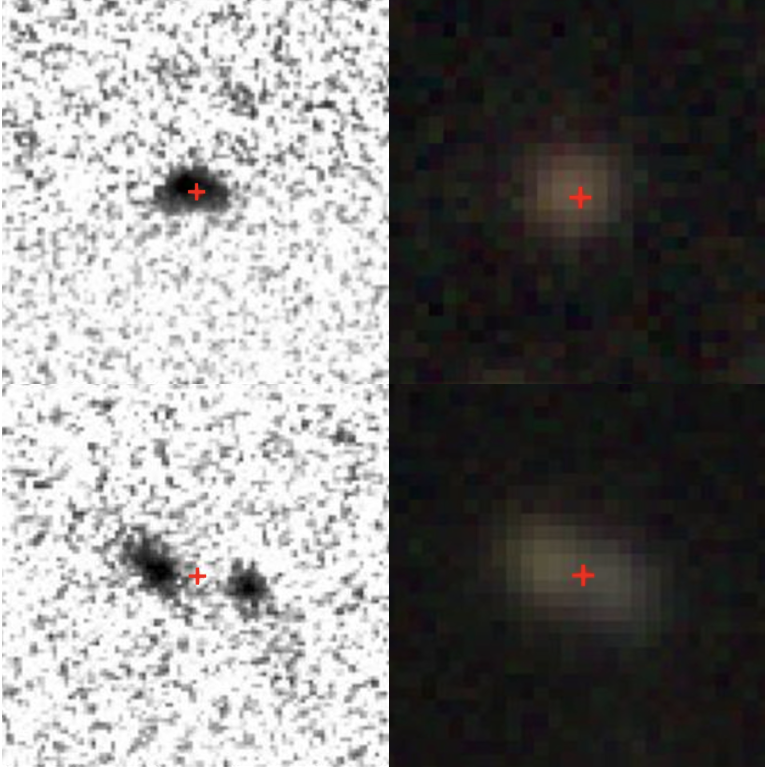}
\caption{\textbf{Top row:} An example of a galaxy where a single source is seen both in the HST (left) and HSC (right) images. The HSC centroid is shown using a red cross. \textbf{Bottom row:} An example of a system where the HSC centroid lies between two objects in the HST image. These objects are blended in the HSC image. The colour of the pixels in the HSC image on either side of the (blended) object are slightly different, indicating that the HSC object is indeed likely to be a blend of two separate objects.}
\label{fig:blended_example}
\end{figure}

As noted in Section \ref{sec:visualinspection}, we use the higher resolution HST images to identify cases where the HSC photometry may have contributions from two separate sources. Figure \ref{fig:blended_example} shows an example of a galaxy where a single source is seen both in the HSC and HST images (top row) and an example of a system where the HSC centroid (shown using the red cross) lies between two objects in the HST image. These objects are blended in the HSC image. The colours of the pixels in the HSC image on either side of the blended object are slightly different, indicating that the HSC object is indeed likely to be a blend of two separate objects (as indicated by the HST image at this location). We remove objects similar to this one from our analysis.  

\bibliographystyle{mnras}
\bibliography{paper} 

\bsp	
\label{lastpage}
\end{document}